\documentclass[11pt,amsmath,amssymb,nofootinbib,aps]{revtex4}

\begin{document}

\title{Constraints on spacetime anisotropy and Lorentz violation from the GRAAL experiment}

\author{Zhe Chang$^{1,2}$\footnote{E-mail: changz@ihep.ac.cn}}
\author{Sai Wang$^{1}$\footnote{E-mail: wangsai@ihep.ac.cn}\footnote{E-mail: saiwangihep@gmail.com}
\footnote{Corresponding author at IHEP, CAS, 100049 Beijing, China.}}
\affiliation{${}^1$\small{Institute of High Energy Physics\\Chinese Academy of Sciences, 100049 Beijing, China}\\
${}^2$\small{Theoretical Physics Center for Science Facilities\\ Chinese Academy of Sciences, 100049 Beijing, China}}

\begin{abstract}
The GRAAL experiment could constrain the variations of the speed of light. The anisotropy of the speed of light may imply that the spacetime is anisotropic. Finsler geometry is a reasonable candidate to deal with the spacetime anisotropy. In this paper, the Lorentz invariance violation (LIV) of the photon sector is investigated in the locally Minkowski spacetime. The locally Minkowski spacetime is a class of flat Finsler spacetime and refers a metric with the anisotropic departure from the Minkowski one. The LIV matrices used to fit the experimental data are represented in terms of these metric deviations. The GRAAL experiment constrains the spacetime anisotropy to be less than \(10^{-14}\). In addition, we find that the simplest Finslerian photon sector could be viewed as a geometric representation of the photon sector in the minimal standard model extension (SME).
\end{abstract}
\maketitle

\section{Introduction}
In the standard model, the speed of light \(c\) is an isotropic constant in the vacuum.
This principle displays one cornerstone for Einstein's special relativity (SR).
Various observations have been proposed to test its validity,
see, for example, the Ref.~\cite{Data tables for Lorentz and CPT violation} (and references therein).
The pioneer and most famous one is the so-called Michelson-Morley experiment \cite{Michelson-Morley1887}.
However, the Michelson-Morley experiment and its decedents involve the two-way propagation of light.
Thus, they refer to the average speed of light \cite{REVIEW OF ONE-WAY AND TWO-WAY EXPERIMENTS}.
Recently, a one-way experiment was performed by the GRAAL facility in the European Synchrotron Radiation Facility (ESRF)
\cite{MPLA2005,NuovoCimentoB2007,Gurzadyan2010,PRL2010}.
It involves the Compton scattering of the high-energy electrons on the photons.
This is the first kinematical non-threshold approach to test the Lorentz symmetry in the collision physics \cite{PRL2010}.
The one-way experiment is sensitive to the first-order variation of the speed of light \cite{REVIEW OF ONE-WAY AND TWO-WAY EXPERIMENTS}.

In the GRAAL's setup, the high-energy electrons collide head-on with the monochromatic laser photons.
The observable quantity is the minimum energy of scattered electrons.
The minimum energy is called the Compton-edge (CE) energy.
It is related with the CE position of the scattered electrons on the detector.
Thus, the GRAAL experiment is also called the Compton-edge experiment.
The CE energy of the scattered electrons would not change when the speed of light is isotropic as postulated in the SR.
Otherwise, it would vary with the spatial directions due to the Earth's spin (precisely, the sidereal rotation).
Therefore, the azimuthal variation of the CE energy could reveal the anisotropy of the speed of light.

The anisotropy of the speed of light implies the breaking of the Lorentz invariance.
The Lorentz invariance violation (LIV) may stem from the quantum-gravity (QG) effects at very high-energy scale.
String theory is the most promising fundamental theory to study the QG.
It could induce the observable low-energy new physics, such as the spontaneous Lorentz invariance violation (sLIV) \cite{SLIV}.
There is an effective field theory to account for the sLIV effects, namely the standard model extension (SME) \cite{SME01,SME02}.
In the SME, possible terms concerning the LIV are added into the Lagrangian of the standard model (SM) by hand.
Recently, the minimal SME was found to be related with Finsler geometry \cite{SME Bogoslovsky01,SME Bogoslovsky02,Kostelecky_Finsler}.
In addition, there exist other LIV models relating to Finsler geometry (see review in Ref.~\cite{SME and Finsler}).
Thus, the anisotropic speed of light may imply an anisotropic Finsler structure of the spacetime.

The LIV could emerge in the Finsler spacetime naturally.
Irrespective of the string motivated sLIV, Bogoslovsky {\it et al.} \cite{A special-relativistic theory of the locally anisotropic space-time. I,A special-relativistic theory of the locally anisotropic space-time. II,A special-relativistic theory of the locally anisotropic space-time. Appendix} proposed a flat Finsler spacetime, the metric of which is invariant under the eight-parametric inhomogeneous transformation group of the event coordinates.
This eight-parametric group is known as the \(DISIM_{b}(2)\) group in the general very special relativity (VSR) \cite{VSR,VSR in Finsler}.
It refers to the partially broken spatial isotropy.
There is also a family of the flat Finsler event spacetime with entirely broken spatial isotropy \cite{Geometrical Models of the Locally Anisotropic Space-Time}.
This refers to the seven-parametric inhomogeneous transformation group of the relativistic symmetry.
In the two families of the flat Finsler event spacetimes, the LIV occurs while the relativistic symmetry (i.e., the principle of special relativity) preserves.
This is different with that in the SME, where the LIV is accompanied by the violation of the relativistic symmetry.
The above discussions are also applied to the LIV model of the electromagnetic field as should be discussed below.
It can not be excluded that not only the Lorentz symmetry, but also, for example, the \(DISIM_{b}(2)\) symmetry, would be broken either partially or entirely at the Planck scale.
Thus, it is interesting to investigate this possibility in a generic flat Finsler spacetime.

Finsler geometry \cite{Book by Rund,Book by Bao, Book by Shen} gets rid of the quadratic restriction on the spacetime line element.
The propagation of a particle in Finsler spacetime may be governed by a modified dispersion relation (MDR).
The Finsler spacetime is intrinsically anisotropic since it preserves less Killing vectors
than the Riemann one \cite{Finsler isometry by Wang,Finsler isometry LiCM}.
The four dimensional Finsler structure admits no more than eight symmetries than the Riemann structure does \cite{A special-relativistic theory of the locally anisotropic space-time. I,A special-relativistic theory of the locally anisotropic space-time. II,A special-relativistic theory of the locally anisotropic space-time. Appendix}.
Recently, we proposed a LIV model of the electromagnetic field in the
locally Minkowski spacetime \cite{Electromagnetic field model in the flat Finsler spacetime}.
The locally Minkowski spacetime \cite{Book by Bao} is a class of flat Finsler spacetime.
It is a straightforward generalization of the Minkowski spacetime.
The Lagrangian was presented explicitly for the electromagnetic field in such a spacetime.
Formally, it could be related to the photon sector of the SME \cite{SME and Finsler,Electromagnetic field model in the flat Finsler spacetime}.
The LIV effects could be viewed as the influence from an anisotropic media on the electromagnetic field.
It is noteworthy that this LIV model also does not violate the principle of special relativity.

In this paper, we first review shortly the GRAAL experiment as well as its constraints on
the anisotropy of the speed of light and on certain LIV parameters.
In the locally Minkowski spacetime, the LIV matrices used to fit the experimental data are investigated for the photon sector.
They could be represented in terms of the metric deviation of the locally Minkowski spacetime from the Minkowski one.
We demonstrate the relationship and the differences between these LIV matrices and those in the SME.
In addition, the Finslerian analysis of the GRAAL experiment could reveal the deeper relationship between both models.
The rest of the paper is arranged as follows.
In section 2, the GRAAl experiment as well as its constraint on the LIV is reviewed in the phenomenological framework.
In section 3, we investigate the LIV of the photon sector in the locally Minkowski spacetime.
The LIV matrices of the photon sector could be obtained and compared with those of the SME.
The GRAAL experiment would constrain the level of the spacetime anisotropy.
Conclusions and remarks are listed in section 4.

\section{The GRAAL experiment and the anisotropy of the speed of light}
Phenomenologically, we describe the GRAAL experiment as well as its constraints on the anisotropy of the speed of light and the LIV.
More details could be found in Ref.~\cite{MPLA2005,PRL2010,Theoretical diagnosis on light speed anisotropy by Zhou and Ma2010}.
In the GRAAL experiment, the \(6.03~\rm{GeV}\) electrons collide head-on with a beam of monochromatic laser photons.
The 4-momentums of the incoming electrons and photons are given by \(p^{\mu}=(E(p),p\hat{p})\)
and \(\lambda^{\mu}=(\omega,-\lambda\hat{p})\), respectively.
The MDR of a photon is given by
\begin{equation}
\label{Phenomenological MDR of Photons}
w=(1+\delta(-\lambda\hat{p}))\lambda\ ,
\end{equation}
where \(\delta(-\lambda\hat{p})\) depends on the direction \(-\hat{p}\) when the space is anisotropic.
It is related with a modified refractive index \(n(-\lambda\hat{p})=1-\delta(-\lambda\hat{p})\) at first order.
The dispersion relation of electrons remains unchanged \(E(p)=\sqrt{p^{2}+m_{e}^{2}}\).
This equals to a choice of the frame such that one measures the speed of light relative to the electrons \cite{PRL2010}.

The CE energy of electrons could be obtained when the scattered photons follow the direction \(\hat{p}\).
In general, the net 4-momentum of a scattering process is assumed to be conserved even though the Lorentz symmetry is broken.
For the CE scattering process, the 4-momentum conservation implies
\begin{eqnarray}
\label{4-momentum conservation equation}
E(p)+\omega &=& E(p')+\omega'\ ,\\
p\hat{p}-\lambda\hat{p} &=& p'\hat{p}+\lambda'_{CE}\hat{p}\ ,
\end{eqnarray}
where the primes denote the 4-momentum of the outgoing electrons and photons.
At the first order, we could obtain the CE energy of photons,
\begin{equation}
\label{CE energy of photons}
\lambda'_{CE}=\lambda_{CE}\left(1+\frac{2\gamma^{2}}{1+{4\gamma\lambda}/{m_{e}}}\delta(\lambda'\hat{p})\right)\ ,
\end{equation}
where \(\gamma={E(p)}/{m_{e}}\) denotes the Lorentz factor of photons
and the CE energy is \(\lambda_{CE}={4\gamma^{2}\lambda}/{(1+{4\gamma\lambda}/{m_{e}})}\) in the SR.
Thus, the CE energy \(\lambda'_{CE}\) would vary azimuthally due to the sidereal rotation of the Earth
when the speed of light is anisotropic in the space.

In the minimal SME, the Lorentz-breaking Lagrangian of the pure photon sector is given by \cite{SME02}
\begin{equation}
\label{Lagrangian in SME}
\mathcal{L}_{photon}=-\frac{1}{4}\eta^{\mu\rho}\eta^{\nu\sigma}F_{\mu\nu}F_{\rho\sigma}
+\mathcal{L}^{CPT-even}_{photon}+\mathcal{L}^{CPT-odd}_{photon}\ ,
\end{equation}
where
\begin{eqnarray}
\label{CPT-even LIV Lagrangian at first order in SME}
\mathcal{L}^{CPT-even}_{photon}&=&-\frac{1}{4}(k_{F})^{\mu\nu\rho\sigma}F_{\mu\nu}F_{\rho\sigma}\ ,\\
\label{CPT-odd LIV Lagrangian at first order in SME}
\mathcal{L}^{CPT-odd}_{photon}&=&\frac{1}{2}(k_{AF})_{\alpha}\epsilon^{\alpha\beta\mu\nu}A_{\beta}F_{\mu\nu}\ .
\end{eqnarray}
The coefficient \(k_{F}\) has the symmetry of the Riemannian tensor and is doubly traceless.

The LIV parameters commonly used to fit the experimental data belong to a decomposition of the \((k_{F})^{\mu\nu\rho\sigma}\) coefficient:
\((\tilde{\kappa}_{e+})^{jk}\), \((\tilde{\kappa}_{e-})^{jk}\), \((\tilde{\kappa}_{o+})^{jk}\), \((\tilde{\kappa}_{o-})^{jk}\)
and \(\tilde{\kappa}_{tr}\).
Here we discard the \(k_{AF}\) term since this term vanishes in the Finslerian photon sector
as should be discussed in the following section.
These LIV matrices are given as follows \cite{Signals for Lorentz violation in electrodynamics}
\begin{eqnarray}
\label{decompose of k}
(\tilde{\kappa}_{e+})^{jk}&=&-(k_{F})^{0j0k}+\frac{1}{4}\epsilon^{jpq}\epsilon^{krs}(k_{F})^{pqrs}\ ,\nonumber\\
(\tilde{\kappa}_{e-})^{jk}&=&-(k_{F})^{0j0k}-\frac{1}{4}\epsilon^{jpq}\epsilon^{krs}(k_{F})^{pqrs}+\frac{2}{3}(k_{F})^{0l0l}\delta^{jk}\ ,\nonumber\\
(\tilde{\kappa}_{o+})^{jk}&=&-\frac{1}{2}\epsilon^{jpq}(k_{F})^{0kpq}+\frac{1}{2}\epsilon^{kpq}(k_{F})^{0jpq}\ ,\nonumber\\
(\tilde{\kappa}_{o-})^{jk}&=&\frac{1}{2}\epsilon^{jpq}(k_{F})^{0kpq}+\frac{1}{2}\epsilon^{kpq}(k_{F})^{0jpq}\ ,\nonumber\\
\tilde{\kappa}_{tr}&=&-\frac{2}{3}(k_{F})^{0j0j}\ .
\end{eqnarray}
The first four matrices are traceless \(3\times3\) matrices while the last one is a single coefficient.
The third one is anti-symmetric while others are symmetric.
Their components could be seen from the left two columns in the Table~(\ref{decomposition}).
\begin{table*}
\begin{center}
\caption{The decomposition (\ref{decompose of k}) of the coefficient \(k_{F}\) for the photon sector as well as its ``formal'' relation with the Finslerian metric deviation \(h\). The first column denotes the components of all LIV observables. These components are defined in the second column. Their corresponding representations in the locally Minkowski spacetime are listed in the third column.}
\label{decomposition}
\begin{tabular}{ccc}
  \hline\hline
  Symbol & SME definitions & Finslerian model \\
  \hline
  \((\tilde{\kappa}_{e+})^{12}\) & \(-(k_{F})^{0102}+(k_{F})^{2331}\) & \(0\)  \\
  \((\tilde{\kappa}_{e+})^{13}\) & \(-(k_{F})^{0103}+(k_{F})^{2312}\) & \(0\)  \\
  \((\tilde{\kappa}_{e+})^{23}\) & \(-(k_{F})^{0203}+(k_{F})^{3112}\) & \(0\)  \\
  \((\tilde{\kappa}_{e+})^{11}-(\tilde{\kappa}_{e+})^{22}\) &~~~ \(-(k_{F})^{0101}+(k_{F})^{2323}+(k_{F})^{0202}-(k_{F})^{3131}\) & \(0\)  \\
  \((\tilde{\kappa}_{e+})^{33}\) & \(-(k_{F})^{0303}+(k_{F})^{1212}\) & \(\frac{1}{2}(h^{00}-h^{11}-h^{22}-h^{33})\)  \\\\
  \((\tilde{\kappa}_{e-})^{12}\) & \(-(k_{F})^{0102}-(k_{F})^{2331}+\frac{2}{3}(k_{F})^{0l0l}\) & \(-h^{12}-\tilde{\kappa}_{tr}\)  \\
  \((\tilde{\kappa}_{e-})^{13}\) & \(-(k_{F})^{0103}-(k_{F})^{2312}+\frac{2}{3}(k_{F})^{0l0l}\) & \(-h^{13}-\tilde{\kappa}_{tr}\)  \\
  \((\tilde{\kappa}_{e-})^{23}\) & \(-(k_{F})^{0203}-(k_{F})^{3112}+\frac{2}{3}(k_{F})^{0l0l}\) & \(-h^{23}-\tilde{\kappa}_{tr}\)  \\
  \((\tilde{\kappa}_{e-})^{11}-(\tilde{\kappa}_{e-})^{22}\) &~~~ \(-(k_{F})^{0101}-(k_{F})^{2323}+(k_{F})^{0202}+(k_{F})^{3131}\) & \(h^{22}-h^{11}\) \\
  \((\tilde{\kappa}_{e-})^{33}\) & \(-(k_{F})^{0303}-(k_{F})^{1212}+\frac{2}{3}(k_{F})^{0l0l}\) & \(\frac{1}{2}(h^{00}+h^{11}+h^{22}-h^{33})-\tilde{\kappa}_{tr}\)  \\\\
  \((\tilde{\kappa}_{o+})^{12}\) & \((k_{F})^{0131}-(k_{F})^{0223}\) & \(-h^{03}\)  \\
  \((\tilde{\kappa}_{o+})^{31}\) & \((k_{F})^{0323}-(k_{F})^{0112}\) & \(-h^{02}\)  \\
  \((\tilde{\kappa}_{o+})^{23}\) & \((k_{F})^{0212}-(k_{F})^{0331}\) & \(-h^{01}\)  \\\\
  \((\tilde{\kappa}_{o-})^{12}\) & \((k_{F})^{0131}+(k_{F})^{0223}\) & \(0\)  \\
  \((\tilde{\kappa}_{o-})^{13}\) & \((k_{F})^{0112}+(k_{F})^{0323}\) & \(0\)  \\
  \((\tilde{\kappa}_{o-})^{23}\) & \((k_{F})^{0212}+(k_{F})^{0331}\) & \(0\)  \\
  \((\tilde{\kappa}_{o-})^{11}-(\tilde{\kappa}_{o-})^{22}\) &~~~ \(2(k_{F})^{0123}-2(k_{F})^{0231}\) & \(0\)  \\
  \((\tilde{\kappa}_{o-})^{33}\) & \(2(k_{F})^{0312}\) & \(0\)  \\\\
  \(\tilde{\kappa}_{tr}\) & \(-\frac{2}{3}(k_{F})^{0l0l}\) & \(\frac{2}{3}h^{00}+\frac{1}{3}(h^{00}-h^{11}-h^{22}-h^{33})\)  \\
  \hline\hline
\end{tabular}
\end{center}
\end{table*}

The GRAAL experiment has been used to constrain the LIV matrix \(\tilde{\kappa}_{o+}\) \cite{PRL2010}.
In the minimal SME, the parameter \(\delta\) in Eq.~(\ref{Phenomenological MDR of Photons}) is given by
\begin{equation}
\label{delta in SME}
\delta(-\lambda\hat{p})=\overrightarrow{\kappa}\cdot\hat{p}\ ,
\end{equation}
where \(\overrightarrow{\kappa}\) denotes \(((\tilde{\kappa}_{o+})^{23},(\tilde{\kappa}_{o+})^{31},(\tilde{\kappa}_{o+})^{12})\).
The GRAAL setup is given as \(\hat{p^{i}}(t)\simeq(0.9~\rm{cos}(\Omega t),~0.9~\rm{sin}(\Omega t),~0.4)\), \(p=6.03~\rm{GeV}\), \(\lambda=3.5~\rm{eV}\), and \(\Omega\simeq 2\pi/(23~\rm{h}~56~\rm{min})\) about the spin axis of the Earth.
By substituting (\ref{delta in SME}) into (\ref{CE energy of photons}), we obtain the CE energy of photons as
\begin{equation}
\label{CE in SME}
\lambda'_{CE}=\tilde{\lambda}_{CE}+0.9\frac{2\gamma^{2}\lambda_{CE}}{1+4\gamma\lambda/m_{e}}
\sqrt{\left((\tilde{\kappa}_{o+})^{23}\right)^{2}+\left((\tilde{\kappa}_{o+})^{31}\right)^{2}}\rm{sin}{\Omega t}\ .
\end{equation}
Here the time-independent \((\tilde{\kappa}_{o+})^{12}\) term is absorbed into \(\tilde{\lambda}_{CE}\) and an irrelevant phase is disregarded.
The above equation (\ref{CE in SME}) reveals that the anisotropy of the speed of light is characterized
by \((\tilde{\kappa}_{o+})^{23}\) and \((\tilde{\kappa}_{o+})^{31}\) in the GRAAL experiment.

The GRAAL experiment found no evidence for the sidereal variation of the CE energy.
This provides an upper bound \cite{PRL2010}
\begin{equation}
\label{deltalambdaCE}
\frac{\Delta\lambda_{CE}}{\lambda_{CE}}<2.5\times10^{-6}~~\rm{(95\%~C.L.)}\ .
\end{equation}
Thus, the LIV parameter acquires an upper limit
\begin{equation}
\sqrt{\left((\tilde{\kappa}_{o+})^{23}\right)^{2}+\left((\tilde{\kappa}_{o+})^{31}\right)^{2}}<1.6\times10^{-14}~~\rm{(95\%~C.L.)}\ .
\end{equation}
In addition, the GRAAL experiment could constrain the LIV matrix of photons in
the standard model supplement (SMS) \cite{Theoretical diagnosis on light speed anisotropy by Zhou and Ma2010}.
In the following section, we would show that it could also constrain the LIV matrices in the Finslerian photon sector.

\section{The photon sector and the spacetime anisotropy}
Finsler geometry is a reasonable candidate to investigate the spacetime anisotropy.
The Finsler spacetime stems from the integral of the form \(s=\int_{a}^{b}F(x,y)d\tau\)
where \(y=dx/d\tau\) denotes a 4-momentum of a particle and \(x\) a location.
The integrand \(F(x,y)\) is positively homogeneous of degree one on \(y\).
The Finsler metric is defined as \(g_{\mu\nu}(x,y)=\frac{\partial}{\partial y^{\mu}}\frac{\partial}{\partial y^{\nu}}\left(\frac{1}{2}F^{2}\right)\) \cite{Book by Bao}.
The locally Minkowski spacetime has a Finsler metric of the form \(g_{\mu\nu}(x,y)=g_{\mu\nu}(y)\).
It is a class of flat Finsler spacetime.
Its metric depends on the 4-momentum \(y\) only.
This is different from the Minkowski metric \(\eta_{\mu\nu}\) which is independent of \(x\) and \(y\).
Thus, the locally Minkowski metric may acquires certain corrections from the new physics at high-energy scales.
For detailed discussions on Finsler geometry, see Ref.~\cite{Book by Rund,Book by Bao,Book by Shen}.

The Finsler spacetime is intrinsically anisotropic.
It could contain more complicated P and T properties than the Riemann spacetime.
Its metric deviation from the Riemann metric may be even, odd or hybrid under the CPT \cite{SME and Finsler}.
For instance, the deviation of the locally Minkowskian Randers metric is hybrid under the CPT \cite{SME and Finsler}.
The Randers metric deviation is given by \(g^{\mu\nu}-\eta^{\mu\nu}=b^{\mu}b^{\nu}+\frac{\beta}{\alpha}\left(\eta^{\mu\nu}-\frac{y^{\mu}y^{\nu}}{\alpha^{2}}\right)
+\frac{1}{\alpha}\left(b^{\mu}y^{\nu}+b^{\nu}y^{\mu}\right)\) where \(\alpha=\sqrt{\eta_{\mu\nu}y^{\mu}y^{\nu}}\) and \(\beta=b_{\mu}y^{\mu}\).
Under the CPT, \(\beta\) changes its sign while \(\alpha\) does not.
Thus, the first term in the right-hand side of the Randers metric deviation remains unchanged
while the last two terms change their signs under the CPT transformation.
This example reveals that the asymmetry of Finsler spacetime leads to the possible LIV and CPT violation (CPTV).

In the recent work \cite{Electromagnetic field model in the flat Finsler spacetime},
a LIV model of the electromagnetic field was proposed in the locally Minkowski spacetime.
The LIV effects originate from the replacement of the Minkowski metric with the locally Minkowski metric
in the Lagrangian of the electromagnetic field.
The Lorentz-breaking Lagrangian of the photon sector is given by
\begin{equation}
\label{Lagrangian}
\mathcal{L}=-\frac{1}{4}g^{\mu\rho}g^{\nu\sigma}F_{\mu\nu}F_{\rho\sigma}=
-\frac{1}{8}(g^{\mu\rho}g^{\nu\sigma}-g^{\nu\rho}g^{\mu\sigma})F_{\mu\nu}F_{\rho\sigma}\ ,
\end{equation}
where \(F_{\mu\nu}=\partial_{\mu}A_{\nu}-\partial_{\nu}A_{\mu}\) denotes the electromagnetic field strength
and \(g^{\mu\nu}\) denotes the locally Minkowski metric.
In the last step, we had anti-symmetrized the spacetime indices \(\mu\nu\) and \(\rho\sigma\).

To study the LIV phenomenologies, we represented the above Lagrangian of the photon sector in the framework of the effective field theory.
First, we expanded the locally Minkowski metric to the first order
\begin{equation}
\label{metric expansion}
g^{\mu\nu}=\eta^{\mu\nu}+h^{\mu\nu}+\mathcal{O}(h^{2})\ ,
\end{equation}
where \(h^{\mu\nu}\) dentes the leading-order departure of the locally Minkowski metric from the Minkowski metric.
The metric deviation \(h^{\mu\nu}\) characterizes all the possible LIV and CPTV of the photon sector.
At the first order of the LIV, the Lagrangian (\ref{Lagrangian}) could be rewritten as
\begin{equation}
\label{Lagrangian at first order}
\mathcal{L}=-\frac{1}{4}\eta^{\mu\rho}\eta^{\nu\sigma}F_{\mu\nu}F_{\rho\sigma}+\mathcal{L}_{LIV}\ ,
\end{equation}
where
\begin{eqnarray}
\label{LIV Lagrangian}
\mathcal{L}_{LIV}=&-&\frac{1}{8}(\eta^{\mu\rho}h^{\nu\sigma}-\eta^{\nu\rho}h^{\mu\sigma}-\eta^{\mu\sigma}h^{\nu\rho}
+\eta^{\nu\sigma}h^{\mu\rho})\nonumber\\ &\times& F_{\mu\nu}F_{\rho\sigma}\ .
\end{eqnarray}
The first term in the right-hand side of Eq.~(\ref{Lagrangian at first order}) denotes the Lorentz invariant part of the Lagrangian
while the second term (namely Eq.~(\ref{LIV Lagrangian})) involves all the possible LIV effects in the photon sector.

The results (\ref{Lagrangian at first order}) and (\ref{LIV Lagrangian}) could be compared with those of the SME.
We could obtain two ``formal'' relations
\begin{eqnarray}
\label{relation 1}
(k_{F})^{\mu\nu\rho\sigma}&=&\frac{1}{2}\left(\eta^{\mu\rho}h^{\nu\sigma}-\eta^{\nu\rho}h^{\mu\sigma}-\eta^{\mu\sigma}h^{\nu\rho}
+\eta^{\nu\sigma}h^{\mu\rho}\right)\ ,\\
\label{relation 2}
(k_{AF})_{\alpha}&=&0\ .
\end{eqnarray}
Note that they are formal relations between the LIV parameters of the SME photon sector and those of the Finslerian one.
One reason is that \(h^{\mu\nu}\) could have complicated CPT property.
The right-hand side of Eq.~(\ref{relation 1}) would have corresponding CPT properties.
As is mentioned above, for instance, the Randers metric deviation \(h\) consists of the CPT-even and CPT-odd parts.
It is CPT-hybrid and the corresponded \(k_{F}\)-components become CPT-hybrid.
However, the coefficient \(k_{F}\) are CPT-even in the SME photon sector.
The other reason involves the number of the independent components of \(k_{F}\).
There are nineteen for the SME photon sector while just ten for the Finslerian one.
The non-vanishing components of \(k_{F}\) are listed in the Table~(\ref{k}) for the Finslerian photon sector.
\begin{table}
\centering
\caption{The formal relation~(\ref{relation 1}) between the coefficient \(k_{F}\) and the Finsler metric deviation \(h\). The components of \(k_{F}\) listed are non-vanishing (their symmetrized and anti-symmetrized counterparts are discarded here). We note that only ten of them are independent for the Finslerian photon sector. This is different from that there are nineteen independent components in the SME photon sector.}
\label{k}
\begin{tabular}{c|c}
  \hline\hline
  Components & Finsler LIV parameters \\
  \hline
  \((k_{F})^{0101}\) & \(\frac{1}{2}(h^{11}-h^{00})\) \\
  \((k_{F})^{0102}\) & \(\frac{1}{2}h^{12}\) \\
  \((k_{F})^{0103}\) & \(\frac{1}{2}h^{13}\) \\
  \((k_{F})^{0112}\) & \(\frac{1}{2}h^{02}\) \\
  \((k_{F})^{0113}\) & \(\frac{1}{2}h^{03}\) \\
  \((k_{F})^{0202}\) & \(\frac{1}{2}(h^{22}-h^{00})\) \\
  \((k_{F})^{0203}\) & \(\frac{1}{2}h^{23}\) \\
  \((k_{F})^{0212}\) & \(-\frac{1}{2}h^{01}\) \\
  \((k_{F})^{0223}\) & \(\frac{1}{2}h^{03}\) \\
  \((k_{F})^{0303}\) & \(\frac{1}{2}(h^{33}-h^{00})\) \\
  \((k_{F})^{0313}\) & \(-\frac{1}{2}h^{01}\) \\
  \((k_{F})^{0323}\) & \(-\frac{1}{2}h^{02}\) \\
  \((k_{F})^{1212}\) & \(-\frac{1}{2}(h^{11}+h^{22})\) \\
  \((k_{F})^{1213}\) & \(-\frac{1}{2}h^{23}\) \\
  \((k_{F})^{1223}\) & \(\frac{1}{2}h^{13}\) \\
  \((k_{F})^{1313}\) & \(-\frac{1}{2}(h^{11}+h^{33})\) \\
  \((k_{F})^{1323}\) & \(-\frac{1}{2}h^{12}\) \\
  \((k_{F})^{2323}\) & \(-\frac{1}{2}(h^{22}+h^{33})\) \\
  \hline\hline
\end{tabular}
\end{table}

According to Eq.~(\ref{decompose of k}) and Eq.~(\ref{relation 1}),
the LIV matrices \(\tilde{\kappa}\) used to fit the experimental data could be represented in terms of the Finslerian metric deviation \(h\).
The obtained results are listed in the third column of Table~(\ref{decomposition}).
In this way, the LIV observables are showed to relate with the anisotropy of the locally Minkowski spacetime.
Once again, we find that only ten LIV parameters remain independent for the Finslerian photon sector.
Especially, we find the representations \((\tilde{\kappa}_{o+})^{23}=-h^{01}\) and \((\tilde{\kappa}_{o+})^{31}=-h^{02}\)
which are closely related with the GRAAL experiment as should be discussed in the following.

The 4-momentum of a particle is conserved in the locally Minkowski spacetime.
This result could be deduced from the isometric transformations \cite{Finsler isometry LiCM,no Cherenkov in Finsler,FSR}.
At the first order, the MDR of a photon was given by \cite{Electromagnetic field model in the flat Finsler spacetime}
\begin{equation}
\label{MDR of photon}
g_{\mu\nu}\lambda^{\mu}\lambda^{\nu}=\omega^{2}-\lambda^{2}+2h_{0i}\omega\lambda^{i}=0\ ,
\end{equation}
where \(\omega^{2}=g_{00}\lambda^{0}\lambda^{0}\) and \(\lambda^{2}=-g_{ij}\lambda^{i}\lambda^{j}\).
It could be rewritten as
\begin{equation}
\label{MDR}
\omega\simeq(1-h_{0i}(\lambda^{j})\hat{\lambda^{i}})\lambda\ ,
\end{equation}
where \(\hat{\lambda^{i}}\) denotes a unit 3-vector direction of the photon.
Here we have represented \(h\) as a function of 3-momentum \(\lambda^{j}\) of the photon.
In the following, we just consider the simplest case that \(h_{0i}(\lambda^{j})\) is a function of only \(\lambda\),
namely the magnitude of the 3-momentum \(\lambda^{j}\).
The metric deviation \(h_{0i}(\lambda)\) becomes constant once the GRAAL setup is given.

According to Eq.~(\ref{Phenomenological MDR of Photons}) and Eq.~(\ref{CE energy of photons}),
the Eq.~(\ref{MDR}) implies the CE energy of the outgoing photons as
\begin{equation}
\label{final CE of photons}
\lambda'_{CE}=\tilde{\lambda}_{CE}+0.9\frac{2\gamma^{2}\lambda_{CE}}{1+4\gamma\lambda/m_{e}}\sqrt{(h_{01})^{2}+(h_{02})^{2}}\rm{sin}{\Omega t}\ ,
\end{equation}
where one time-independent term is absorbed into \(\tilde{\lambda}_{CE}\) and an irrelevant phase is disregarded.
This reveals that the anisotropy of the speed of light is characterized by the anisotropic parameters \(h_{01}\) and \(h_{02}\)
for the GRAAL experiment in the Finslerian electromagnetic model.
Recall that the parameters \(h_{01}\) and \(h_{02}\) are related to the LIV parameters
\((\tilde{\kappa}_{o+})^{23}\) and \((\tilde{\kappa}_{o+})^{31}\), respectively.
Exactly, the CE energy (\ref{final CE of photons}) would take the same form as in Eq.~(\ref{CE in SME}).
This result reveals that the simplest Finslerian photon sector may be a geometric representation of the SME one.
This coincides with the proposition in Ref.~\cite{SME Bogoslovsky01,SME Bogoslovsky02,Kostelecky_Finsler} that the SME could be related with Finsler geometry.
According to Eq.~(\ref{deltalambdaCE}), a constraint
\begin{equation}
\sqrt{(h_{01})^{2}+(h_{02})^{2}}<1.6\times10^{-14}~\rm{(95\%~C.L.)}
\end{equation}
is set on the anisotropic departure of the locally Minkowski spacetime from the Minkowski one.
It is noteworthy that the ``eather drift'' experiment has provided an upper constraint \(10^{-10}\)
on the spacetime anisotropy \cite{Eather01,Eather02}.
The GRAAL result improves the sensitivity by four orders of magnitude.
However, one should note that both constraints are model-dependent.
In particular, the upper limit \(10^{-10}\) has been obtained within the framework of
the relativistically invariant Finslerian theory \cite{A special-relativistic theory of the locally anisotropic space-time. I,A special-relativistic theory of the locally anisotropic space-time. II,A special-relativistic theory of the locally anisotropic space-time. Appendix},
where the speed of light does not depend on the direction of light and equals to \(c\).

\section{Conclusions and remarks}
The GRAAL experiment takes advantage of the Compton process of the high-energy electrons scattering the laser photons.
The CE energy would vary azimuthally due to the sidereal rotation of the Earth when the speed of light is anisotropic.
The anisotropy of the speed of light could be accounted by the LIV effects.
We followed the convention of decomposing the LIV coefficient \(k_{F}\) of the photon sector into nineteen LIV matrices \((\tilde{\kappa}_{e+})^{jk}\), \((\tilde{\kappa}_{e-})^{jk}\), \((\tilde{\kappa}_{o+})^{jk}\), \((\tilde{\kappa}_{o-})^{jk}\) and \(\tilde{\kappa}_{tr}\).
However, the GRAAL result showed no evidence for the variations of the CE energy.
It gave upper limits on the anisotropy of the speed of light and certain LIV parameters.

The locally Minkowski spacetime is a class of flat Finsler spacetime.
It is a simplest anisotropic flat spacetime and may be viewed as certain modification of the Minkowski spacetime
at the ultra high-energy physics.
It naturally leads to the LIV and the CPTV.
We had proposed a model of the electromagnetic field in the locally Minkowski spacetime in a previous paper.
The LIV coefficient \(k_{F}\) of the photon sector could be formally related to
the anisotropic departure of the Finsler metric from the Minkowski one, while the coefficient \(k_{AF}\) vanishes.
We listed the Table~(\ref{k}) in detail to present the non-vanishing components of \(k_{F}\).
There are only ten independent components for the Finslerian photon sector.
They may have the complicated CPT property.
This is very different from that \(k_{F}\) is CPT-even in the SME.

Formally, we obtained all the LIV matrices \(\tilde{\kappa}\) in terms of
the anisotropic deviation \(h\) of the locally Minkowski metric from the Minkowski one.
We listed these LIV matrices as well as their representation in terms of \(h\) in the Table~(\ref{decomposition}).
In the SME, there are nineteen independent components for this set of LIV matrices.
In the Finslerian photon sector, however, we found that only ten of them are independent.
This prediction reveals that the number of the LIV parameters is severely squeezed for the Finslerian photon sector.
Actually, there are only ten independent components for the anisotropic metric deviation \(h\) in the locally Minkowski spacetime.
Thus, they should completely determine the LIV and CPTV parameters as well as the anisotropy of the speed of light.
As the SME, one can have LIV models without the CPT violation, or reciprocally have the CPT violation in a Lorentz covariant theory \cite{Chaichian2012a,Chaichian2012b,Chaichian2013}.

The discussion on the GRAAL experiment could reveal the deep relationship between the Finsler spacetime and the SME.
In the simplest case, the formula of the CE energy in the Finslerian photon sector was found to be as same as the one in the minimal SME.
The LIV could be characterized by two common parameters in both models.
Thus, the Finslerian photon sector might be a geometric representation of the photon sector in the SME.
This prediction is compatible with the proposition that the SME could be Finsler geometric in
Ref.~\cite{SME Bogoslovsky01,SME Bogoslovsky02,Kostelecky_Finsler}.
In addition, the level of the spacetime anisotropy was constrained to be less than \(10^{-14}\) by the GRAAL experiment.
This constraint improves the one from the previous ``eather drift'' experiment by four orders of magnitude.

\vspace{0.3 cm}

\begin{acknowledgments}
We thank useful discussions with Yunguo Jiang, Danning Li, Ming-Hua Li, Xin Li and Hai-Nan Lin.
This work is supported by the National Natural Science Fund of China under Grant No. 11075166.
\end{acknowledgments}

\end{document}